# Advection-Dominated Models of Luminous Accreting Black Holes

Ramesh Narayan

Harvard-Smithsonian Center for Astrophysics
60 Garden Street, Cambridge, MA 02138

## Abstract

It has been found that a class of optically-thin two-temperature advection-dominated accretion solutions explains many observations of low-luminosity accreting black holes. Here it is shown that these models give a satisfactory description also of higher luminosity systems, provided the viscosity parameter $\alpha$ is large. The models reproduce the spectra of black hole X-ray binaries in the Low State, and explain the transition from the Low State to the High State at a critical mass accretion rate. The models also show that the X-ray/$\gamma$-ray spectra of X-ray binaries and active galactic nuclei should be similar, as confirmed by observations.

Subject Headings: accretion, accretion disks, black hole physics, galaxies: quasars, stars: binaries, X-rays: stars

## 1. Introduction

Recently, a new class of accretion solutions has been identified which is present at relatively low mass accretion rates, $\dot{m} \lesssim 0.3\alpha^2$ (Narayan & Yi 1994, 1995a,b, Abramowicz et al. 1995, Chen 1995, Chen et al. 1995, see Narayan 1995 for a review). Here $\alpha$ is the standard viscosity parameter (Shakura & Sunyaev 1973) and $\dot{m}$ is expressed in Eddington units, $\dot{m} = \dot{M}/\dot{M}_{\rm Edd}$, where $\dot{M}_{\rm Edd} = 2.2 \times 10^{-8} m~M_\odot{\rm yr}^{-1} = 1.4 \times 10^{18} m~{\rm g\,s}^{-1}$ for an accreting star of mass $mM_\odot$ (a fiducial efficiency of 10% is used, cf. Frank, King & Raine 1992).

The key feature of the new solutions is that the gas is optically thin and radiates inefficiently so that most of the viscously generated energy is advected radially. These advection-dominated solutions have a two-temperature structure (Shapiro, Lightman & Eardley 1976, SLE), where the ions are nearly virial, $T_i \sim 10^{12}{\rm K}/r$ with the radius $r$ expressed in Schwarzschild units, and the electrons are relativistic, $T_e \sim 10^9 - 10^{10}$ K. The high temperature causes the flow to have a nearly spherical morphology (Narayan & Yi 1995a), or perhaps a fat toroidal structure as in the related ion torus model of Rees et al. (1982). Because radial pressure forces are important the flow rotates with a sub-Keplerian angular velocity.

Since the solutions under discussion occur at relatively low $\dot{m}$, and have poor radiative efficiencies (because of their advection-dominated nature), they are especially good for modeling low-luminosity systems. Indeed, the solutions have been used successfully to model Sagittarius A*, the ultra-low-luminosity supermassive black hole at the Galactic Center (Narayan, Yi & Mahadevan 1995), the soft X-ray transients (SXTs), A0620-00, V404 Cyg, Nova Muscae 1991, in their quiescent states (Narayan, McClintock & Yi 1996), and the low-luminosity active galactic nucleus (AGN) in the core of NGC 4258 (Lasota et al. 1995). Also, Fabian & Rees (1995) have suggested that most nearby large ellipticals could have supermassive black holes in their centers, but the holes may be accreting in an advection-dominated mode, which would explain the puzzling lack of significant accretion luminosity in the cores of these galaxies (Fabian & Canizares 1988). The advection-dominated models seem to perform quite well in all these applications. The obvious question now is: can the solutions also be used to model higher luminosity systems?

Most luminous accreting black holes, whether they are in X-ray binaries (XRBs) or AGNs, are strong emitters of X-rays and $\gamma$-rays (Tanaka 1989, Grebenev et al. 1993, Gilfanov et al. 1995, Elvis et al. 1994, Johnson et al. 1994). The soft $\gamma$-ray luminosity exceeds $10^{37}$ ergs s$^{-1}$ in the most luminous XRBs and $10^{44}$ ergs s$^{-1}$ in luminous AGNs. The spectra are very hard, with power-law photon indices $\alpha_N \sim 1.5 - 2.5$ in the photon energy range $E \sim$ few keV to few $\times$ 100 keV. The spectra usually turn over at $E \sim$ few $\times$ 100 keV (Gilfanov et al. 1995, Maisack et al. 1993, Johnson et al. 1994), suggestive of a thermal process, and the observed spectra can usually be fit with a Sunyaev-Titarchuk (1980) thermal Comptonization model with optical depth $\tau_{es} \sim 1-$ few and electron temperature $T_e \sim 10^9$ K. There may also be a reflection component in the spectrum due to scattering off cold gas (Lightman & White 1988, George & Fabian 1991).

These observations, mostly carried out with X-ray and gamma-ray telescopes on GINGA, GRANAT and CGRO, indicate the nearly universal presence of hot optically thin thermal gas in accreting black holes. Although this has been established beyond doubt only in recent years, Thorne & Price (1975) and SLE had already noted the fact in the 1970s, based on early observations of Cyg X-1. Indeed, SLE developed a hot two-temperature accretion model, which was studied extensively by several later authors (e.g. Kusunose & Takahara 1985, 1989, White & Lightman 1989, Wandel & Liang 1991, Melia & Misra 1993). However, the SLE model suffers from a violent thermal instability (Piran 1978, Wandel & Liang 1991), which leads to a serious inconsistency in the model.

The new optically-thin advection-dominated solutions studied in this paper are hotter and more optically thin than the SLE solution, but are nevertheless thermally stable (Abramowicz et al. 1995, Narayan & Yi 1995b, Kato, Abramowicz & Chen 1995). These solutions therefore have precisely the properties needed to fit the observations. We study the models here using somewhat higher mass accretion rates than considered in previous studies, which have tended to concentrate on low-luminosity sources. We calculate model spectra for a range of values of $\dot m$ and carry out a qualitative comparison with observations of high-luminosity accreting black holes. The calculations are described briefly in §2, and the results are summarized in §3. §4 discusses the results.

## 2. The Model

We assume that the accretion flow consists of two zones: (i) an outer zone extending from a large radius $r_{\rm out} = 10^6$ down to a transition radius $r_{\rm tr}$, where the gas is in the form of a standard thin accretion disk (e.g. Shakura & Sunyaev 1973, Frank, King & Raine 1992), and (ii) an inner zone extending from $r = r_{\rm tr}$ down to $r = 3$, where the flow is hot, optically thin and advection-dominated. Two different mechanisms have been proposed whereby cold gas in the outer thin disk could be transformed into the hot inner flow. Meyer & Meyer-Hofmeister (1994) have shown that evaporation by means of conductive heating from a corona is an effective way of truncating low-$\dot m$ accretion disks around white dwarfs (Meyer & Meyer-Hofmeister 1994). It seems likely that the same mechanism can operate also in disks around black holes (Narayan & Yi 1995b). Independently, Honma (1995) has suggested that the inner edge of the outer disk can be heated by turbulent transport of energy outward from the advection-dominated interior, leading to evaporation of the cold gas. Both mechanisms suggest a transition radius $r_{\rm tr} \sim 10^3 - 10^5$ at low mass accretion rates.

We assume that the spectrum of the outer thin disk is blackbody in character. In calculating the emission from this zone we include the luminosity due to local viscous dissipation as well as the additional luminosity due to surface reprocessing of X-ray irradiation from the hot inner flow. For calculating the irradiation we assume that the outer disk is geometrically thin and integrate the irradiating flux from $r = 3$ to $r = r_{\rm tr}$, assuming that the optically thin interior flow is spherically symmetric.

For the inner zone, we model the gas as a two-temperature plasma, where we allow the ions to be much hotter than the electrons and we assume that the coupling between ions and electrons is via Coulomb collisions (Narayan & Yi 1995b). It has been proposed that collective instabilities may possibly force a single temperature (e.g. Phinney 1981, Rees et al. 1982), but the only specific mechanism that has been published is that due to Begelman & Chiueh (1988). The relevant parameters of this mechanism are fairly uncertain and it appears that under many circumstances it does not operate rapidly enough to equilibrate the ion and electron temperatures (see Appendix A of Narayan & Yi 1995b). The two-temperature assumption is therefore not unreasonable for our models.

We assume a constant viscosity parameter $\alpha$ within the hot flow. Given that advection-dominated flows are morphologically and dynamically very different from standard thin disks, the value of $\alpha$ in the hot flow could differ from the usual values $\sim 0.01-0.1$ that have been deduced for thin disks. We assume that there is equipartition between gas pressure and magnetic pressure in the hot gas such that $P_{\rm gas} = \beta P_{\rm tot}$, $P_{\rm mag} = (1-\beta)P_{\rm tot}$, with constant $\beta$; radiation pressure is not expected to be important.

At the high temperatures of interest, photons are emitted primarily by synchrotron and bremsstrahlung processes (Narayan & Yi 1995b). Comptonization is extremely important and we calculate its effect shell by shell from $r = 3$ to $r = r_{\rm tr}$. In calculating the Comptonization we include photons generated locally, photons flowing out from smaller radii, and photons from the outer thin disk which have penetrated into the inner flow. We calculate the Comptonization using the formula given in Jones (1968) which is an excellent approximation when the radiation field and the scattering electrons are isotropic. The formula automatically includes the Klein-Nishina correction as well as saturation when the photons approach $E = 3kT_e$. We apply a gravitational redshift as photons move from one radius to another.

We calculate the density, radial velocity, angular velocity, and the ion and electron temperatures of the inner flow by using the self-similar solution of Narayan & Yi (1994) as a local representation of the solution, but adjusting the advection parameter $f$ at each $r$ self-consistently (cf. Narayan & Yi 1995b, Narayan et al. 1996). Although the self-similar solution is based on a height-averaged set of equations, it has been shown to be an excellent representation even of the nearly spherical configurations found in advection-dominated flows (Narayan & Yi 1995a, see also Begelman & Meier 1982). The numerical calculations presented in Narayan & Yi (1994) show that the existence of a transition zone at $r = r_{\rm tr}$, where the inner flow connects to the outer thin disk, poses no serious difficulties to the viability of the hot flow. The flow makes a smooth transition and approaches the self-similar form a short distance in from the transition radius. Therefore, there is very little error in using the local self-similar solution.

A new feature of the present calculations is that we calculate the gas properties and spectrum simultaneously and self-consistently by using a coupled set of equations which we solve iteratively. Another point to note is that the models with high values of $\dot m$, especially those close to the limiting $\dot m$ of this branch of solutions, are not truly advection-dominated but radiate a fair fraction of the energy released through viscosity. We allow for this by solving self-consistently for the radiative efficiency $f$ as a function of radius.

## 3. Results

Each model is described by five parameters: (i) $\alpha$, (ii) $\beta$, (iii) the black hole mass $m$ in solar units, (iv) the Eddington-scaled mass accretion rate $\dot m$, and (v) the transition radius $r_{\rm tr}$ where the outer thin disk gives way to the inner hot flow. The results are fairly insensitive to $\beta$; all calculations reported here correspond to $\beta = 0.5$. The current models of disk evaporation (Meyer & Meyer-Hofmeister 1994, Honma 1995) are not sufficiently developed yet to give precise estimates of the transition radius $r_{\rm tr}$ in individual cases. We therefore take an empirical approach to estimate $r_{\rm tr}$. By fitting observed optical and UV spectra of several SXTs in quiescence, Narayan et al. (1996) found that $r_{\rm tr} \sim$ few $\times 10^3$ in these systems. Using this as a guide we arbitrarily set $r_{\rm tr} = 10^3$ in the present calculations; tests indicate that the results are not very different if we take $r_{\rm tr} = 10^4$. When $\dot m$ exceeds a critical value $\sim 0.1\alpha^2$, we find that it is not possible for an advection-dominated flow to extend out to radii as large as $10^3$. In these cases we reduce $r_{\rm tr}$ as much as necessary to obtain a solution.

Black hole XRBs are known to have a Low State in which the spectrum extends up to a few hundred keV as a very hard power-law with $\alpha_N \sim 1.5 - 2$ (e.g. Tanaka 1989, Grebenev et al. 1993, van der Klis 1994, Gilfanov et al. 1995). The maximum luminosity observed in this state is $\sim$ few $\times 10^{37}$ ergs s$^{-1}$ (Grebenev et al. 1993) which is a few percent of the Eddington luminosity $L_{\rm Edd}$ for a black hole of mass $m = 10$. In our models, we find that the hot inner flow exists for values of $\dot m$ up to $\sim 0.3\alpha^2$. The maximum luminosity we find in this kind of a flow is $\sim 0.05\alpha^2 L_{\rm Edd}$. Therefore, our hot models can produce the maximum luminosities observed in Low State XRBs only if $\alpha$ is fairly large. The results described below correspond to $\alpha = 1$.

Figure 1 shows a sequence of model spectra of black hole XRBs (black hole mass $m = 10$) where we have varied the Eddington-scaled $\dot m$ from $10^{-3}$ to $10^{-0.5}$. There are several notable features in these spectra:
1. For low values of $\dot m$ in the range $10^{-3} - 10^{-2}$, we see that the spectrum has a peak in the optical and UV due to blackbody emission from the outer thin disk plus a fairly hard X-ray/$\gamma$-ray spectrum due to the hot inner flow. Even though the outer and inner zones have the same $\dot m$ and the gravitational energy released in the inner flow is much greater, nevertheless we see that the luminosity is dominated by the outer disk. This is because the hot inner flow is massively advection-dominated. The bumps in the X-ray spectrum are due to successive Compton scatterings in the extremely optically thin gas. The models with $\dot m = 10^{-3}$ and $10^{-2.5}$ are similar to models described by Narayan et al. (1996) for the quiescent SXTs, A0620-00 and Nova Muscae 1991, while the $\dot m = 10^{-2}$ model is similar to the model of V404 Cyg. Note that the luminosity of the outer disk increases linearly with $\dot m$, whereas the inner disk luminosity varies somewhat more steeply than $\dot m^2$ because of advection (see Fig. 11 of Narayan & Yi 1995b). Note also that these spectra have a significant fraction of their X-ray/$\gamma$-ray luminosity in the energy range from a few keV to a few hundred keV. This prediction (Narayan et al. 1996) can be tested with ASCA, XTE, and other hard X-ray telescopes.

2. As $\dot{m}$ increases and approaches $\sim 10^{-1.5} - 10^{-1}$, the hard X-ray luminosity from the inner flow goes up substantially ($L \sim 10^{36} - 10^{37.5}$ ergs s$^{-1}$) and approaches a few percent of $L_{\rm Edd}$, similar to the highest luminosities seen in the Low State (Grebenev et al. 1993). At these accretion rates, the optical depth approaches unity, and the Comptonization yields a smooth power-law hard X-ray spectrum. The models typically give a spectral index $\alpha_N \sim 1.5 - 2$, which agrees very well with observations. The model spectra cut off at around 200 keV, which is again consistent with observations of several sources (e.g. Gilfanov et al. 1995). Note that the hard X-ray luminosity varies almost as $\dot{m}^3$ at these values of $\dot{m}$, which suggests that small variations in $\dot{m}$ could lead to large variations of the hard X-ray luminosity. Indeed, black hole XRBs display large amplitude variability in the Low State (Grebenev et al. 1993, Gilfanov et al. 1995). This could be produced by much smaller amplitude variations in $\dot{m}$. Another feature of the solutions is that the electron temperature decreases with increasing luminosity. This has been seen in the source GRO J0422+32 (J. Kurfess, private communication). Overall, it appears that the models with $\dot{m} = 10^{-1.5} - 10^{-1}$ correspond quite well to the Low State of black hole XRBs and it is tempting to associate this state of black hole systems to advection-dominated accretion.

3. For $\dot{m} \gtrsim 10^{-1}$, we find that the Compton cooling becomes very large because of the increasing optical depth, and as a result the advection-dominated solution is restricted in its radial extent. In terms of the parameter $r_{\rm tr}$, we find that over the range $10^{-1} < \dot{m} < 10^{-0.5}$, the maximum $r_{\rm tr}$ that is allowed falls rapidly from the assumed value of $r_{\rm tr} = 10^3$ down to $r_{\rm tr} = 3$. The spectrum thus evolves rapidly over this $\dot{m}$ range. The blackbody component due to the outer disk grows in importance and moves into the soft X-ray band. The hard X-ray spectrum at the same time softens, with $\alpha_N$ increasing to values greater than 2. Finally, at $\dot{m} = 10^{-0.5}$, there is no advection-dominated zone at all and the entire flow occurs as a standard thin accretion disk down to $r = 3$. These results are reminiscent of several features seen in the data. Black hole XRBs exhibit a High State (and a Very High State, e.g. van der Klis 1994, Gilfanov et al. 1995) where the spectrum consists of a luminous ultra-soft component. The transition from the Low State to the High State appears to occur rather suddenly when the hard X-ray luminosity is $\sim (10^{-2} - 10^{-1}) \times L_{\rm Edd}$, exactly as in the models shown here.

4. One point to note is that even in the High and Very High States, black hole XRBs have a weak hard tail in their spectra. The spectral slope is relatively steep compared to the Low State, and the luminosity in the tail is a small fraction of that in the ultra-soft component. The hard radiation probably arises from a corona (Narayan 1995) which partially reprocesses soft radiation from the disk (e.g. Haardt & Maraschi 1991). Since our model with $\dot{m} = 10^{-0.5}$ consists of a pure thin disk, it has only a soft component in the spectrum and there is no hard tail.

5. The models make some qualitative predictions of the spectral evolution during outbursts of SXTs. When the source initially goes into outburst, the spectrum will become progressively harder until $\dot{m}$ reaches $\sim 10^{-1}$. Then, with further increase in $\dot{m}$, the outer disk will move in and will fill up the hot cavity in the center, leading to a primarily soft spectrum with perhaps a weak hard tail due to a corona. Later, as the outburst dies down, the soft flux will decay (exponentially) and the hard flux will become more dominant so that at the end of the outburst the spectrum will return to its initial hard state. The observations reported by Lund (1993) on the Nova Muscae 1991 outburst are in good agreement with these predictions; the spectrum was hard both early in the outburst and towards the end, but was dominated by a soft component in between. Further, the ultraviolet spectrum of the source during outburst was consistent with a standard thin disk extending to small radii (Shrader et al. 1993), whereas in quiescence the optical data suggest that the disk is truncated at a large radius (Narayan et al. 1996). This is again consistent with our model. In contrast to Nova Muscae 1991, in the case of V404 Cyg the spectrum was found to remain hard throughout the outburst (Sunyaev et al. 1991). Presumably, in this case, $\dot{m}$ remained below about $10^{-1}$ throughout the outburst and so the hot inner flow was never quenched. Since the peak luminosity of Nova Muscae 1991 and V404 Cyg were comparable, the different outburst characteristics in the two cases suggest that the black hole in Nova Muscae has a lower mass than the hole in V404 Cyg (so that at the same physical $\dot{M}$ the latter source has a smaller $\dot{m}$). Independent mass estimates in the two sources (cf. references in Narayan et al. 1996) are consistent with this explanation.

In view of the success of these models in fitting observations of XRBs, it is of interest to ask if the same models can also explain observations of AGNs. Figure 2 shows a sequence of models with $\alpha = 1$, $\beta = 0.5$, $\dot{m} = 10^{-1}$, $r_{\rm tr} = 10^{1.5}$, and with $m$ ranging from 10 to $10^9$. As the black hole mass increases, the blackbody bump due to the outer disk moves from soft X-rays down to the optical/UV band. This is expected for a thin accretion disk since the effective temperature scales as $m^{-1/4}$ (cf. Frank et al. 1992). The properties of the hot inner flow are, however, essentially independent of $m$. Thus, we find that regardless of $m$ all the models have essentially the same electron temperature and optical depth and produce power-law hard X-ray spectra with cutoffs roughly at $100 - 200$ keV.

The model spectrum we obtain for $m = 10^9$ is quite similar to the optical-to-X-ray spectra of normal quasars cataloged by Elvis et al. (1994) and nicely reproduces the blue bump in the optical band and a flat spectrum in the X-ray band. Similarly, the $m = 10^7$ spectrum resembles hard X-ray observations of Seyfert nuclei (Maisack et al. 1993, Johnson et al. 1994). At infrared and millimeter wavelengths, quasars and Seyferts have more emission than our models indicate, but this is probably due to dust reprocessing in the outer accretion disk/torus (Sanders et al. 1989, Pier & Krolik 1992, 1993) which we have not included in our models.

Another effect we have not included is the reflection effect when hard photons irradiate cold gas (Lightman & White 1988, George & Fabian 1991). Considerable evidence has accumulated that the spectra of Seyferts do have reflection bumps, and it appears that the effective solid angle of the cold gas as viewed by the hot gas is often comparable to $2\pi$. In our model, it is possible to have a reflection bump through irradiation of the cool outer disk by hard radiation from the inner flow. However, unless the outer disk is vertically very thick, or $r_{\rm tr}$ is very small, it is not likely that the effective solid angle will be large. This constraint needs to be explored in more detail.

The spectra shown in Fig. 2 correspond to $\dot{m} = 0.1$. Lasota et al. (1995) have proposed a model for the LINER galaxy NGC 4258 using an accretion rate which is approximately an order of magnitude lower than this. (Note that the definition of the Eddington accretion rate $\dot{M}_{\rm Edd}$ is different in the two papers; in the present paper an efficiency factor of 10% has been used to define $\dot{M}_{\rm Edd}$ whereas Lasota et al. do not include this factor.) The model spectrum of Lasota et al. fits the observed spectral slope of NGC 4258 in the ASCA band quite well.

## 4. Discussion

The results presented in this paper suggest that a two-temperature advection-dominated model has many of the qualitative features needed to explain observations of accreting black holes in XRBs and AGNs. The distinction between the Low State and the High State in XRBs and the transition between the two states is well reproduced by the models. The strong similarity in electron temperatures and X-ray spectra between XRBs and AGNs is also explained.

The calculations reported here are fairly detailed and essentially self-consistent, but suffer from one major simplification. The properties of the gas at each radius are not calculated by solving a full coupled system of radial differential equations, but rather are obtained from local solutions involving algebraic equations. Specifically, the Shakura-Sunyaev (1973) solution is used to describe the outer disk, and the self-similar solution of Narayan & Yi (1994, 1995b) is used for the inner advection-dominated flow. The local approximation is excellent at most radii, but is expected to break down close to the black hole. The flow in this region has been studied by a few authors (e.g. Matsumoto, Kato & Fukue 1985, Chakrabarti 1990) who find that the gas makes a sonic transition (usually at $r \sim$ few) before falling supersonically into the black hole. There may also perhaps be a radial shock in the flow (e.g. Chakrabarti 1990). Because of these effects, there is a range of radii close to the black hole where the flow is likely to differ from the self-similar form. In general, we expect the radial velocity to be larger than the self-similar value and the density to be correspondingly lower. The primary effect of this is likely to be that the limiting $\dot{m}$ up to which the advection-dominated solution survives will be higher than the limit $\sim 0.3\alpha^2$ derived on the basis of the local solution. Turning this around, the value of $\alpha$ needed for agreement with the observations may not have to be as large as unity, as indicated by the calculations presented here, but perhaps a smaller value of $\alpha$ may be sufficient.

Other simplifications in the models presented here should also be mentioned. We have simplified the flow geometry by taking the inner flow to be purely spherical and the outer flow to be a very thin disk. This permits us to treat only one spatial dimension (spherical or cylindrical radius) in each zone of the flow. A more realistic two-dimensional model will complicate the calculations enormously, both in the flow dynamics and radiative transfer, but is worthwhile. Further, in the High State, we have not included a hot corona above the thin disk. As explained in sec. 3, such a component could lead to some hard radiation even at high $\dot{m}$. Finally, in some of our high $\dot{m}$ models, the outer thin disk extends down to small radii and enters a region where radiation pressure dominates and the thin disk becomes unstable (Lightman & Eardley 1974). Time dependent calculations would be of interest to follow the development of this instability.

Detailed models with these improvements will change some of the quantitative details of the results. However, the qualitative features highlighted in this paper are likely to survive. The basic result is that the advection-dominated model is quite promising as a general description of accretion flows around black holes.

One outstanding question is whether it is reasonable to set the viscosity parameter $\alpha$ equal to unity, as our models seem to require. We are forced to make this choice in order to explain luminosities $\sim$ few $\times 10^{37}$ ergs s$^{-1}$ seen in the Low State in some black hole XRBs like Cyg X-1, GX339-4, and 1E1740.1-2942 (Grebenev et al. 1993). Since the maximum luminosity we obtain with our two-temperature solution is $\sim 0.05\alpha^2 L_{\rm Edd}$, the observations require $\alpha \sim 1$ for a reasonable black hole mass of $m = 10$. Magnetic stresses (Hawley, Gammie & Balbus 1995) and hydrodynamic instabilities (due to convection, see Narayan & Yi 1994, 1995a, Kumar, Narayan & Loeb 1996) are two potential sources of viscosity in these solutions, but it is not clear if either process is efficient enough to make $\alpha \sim 1$. However, as explained earlier, the actual value of $\alpha$ that is needed may turn out to be somewhat smaller than unity when a full non-local calculation is carried out. Such more detailed models are urgently needed. As an aside, it should be mentioned that although advection-dominated flows are convectively unstable, they do not force the accreting gas to be isentropic. This is because the rapid radial accretion, coupled with entropy generation via viscous dissipation, can overcome the effects of convective energy transport (see Narayan & Yi 1994).

As a final point we note that, although the advection-dominated flows discussed here are quasi-spherical and have substantial pressure support, they differ greatly from true spherical accretion flows which have been studied for many years (Bondi & Hoyle 1944, Bondi 1952, Shvartsman 1971, Shapiro 1973, Ipser & Price 1977). The critical difference is that the new advection-dominated solutions rotate around the black hole so that these solutions are relevant whenever the gas starts off with non-zero angular momentum at large radius. Indeed, the self-similar nature of the solutions ensures that accreting gas will be "attracted" to these solutions starting from a wide range of outer boundary conditions so long as the radiative efficiency is low (Narayan & Yi 1994). In contrast, pure spherical accretion requires the artificial boundary condition that the accreting gas should have effectively *zero* angular momentum on the outside. Even the slightest amount of initial angular momentum would cause rotational support to become significant at some critical radius, and inside of this radius the flow will be compelled to take up a rotating form. One of the features of the pure spherical accretion flow is that the gas makes a sonic transition at a large radius and flows in supersonically over many $e$-folds of the radius (Bondi 1952). Since pressure support is the only counter to gravity in these flows, the "bottom falls out" in some sense and gravity takes charge and causes the gas to rush into the black hole. In contrast, the rotating advection-dominated flows studied in this paper have significant rotational support. This ensures that the bottom cannot fall out until the gas is quite close to the black hole. Consequently, the gas maintains a nearly self-similar form until close to the black hole and only then does it make a sonic transition. Thus, even though the black hole horizon provides no radial support at all, nevertheless the accretion occurs in a controlled manner, mediated by viscous transport of angular momentum.

This work was supported in part by grant AST 9423209 from the NSF. The author thanks Didier Barret and Insu Yi for useful discussions, and the Institute for Theoretical Physics, Univ. of California, Santa Barbara, for hospitality during the late stages of the work.

**Figure Captions**

Fig. 1. Spectra of black hole XRB models with $\alpha = 1$, $\beta = 0.5$, $m = 10$. The plots show $\nu L_\nu$ versus photon energy $E$. In this representation, a spectrum which is horizontal corresponds to a photon index of $\alpha_N = 2$, while a spectrum which rises toward higher $E$ corresponds to $\alpha_N < 2$. Starting from below, the models have $(\log \dot{m}, \log r_{tr}) = (-3, 3), (-2.5, 3), (-2, 3), (-1.5, 3), (-1.1, 3), (-1, 2.2), (-0.9, 1.5), (-0.8, 1), (-0.7, 0.8), (-0.6, 0.6), (-0.5, 0.5)$.

Fig. 2. Spectra of models with $\alpha = 1$, $\beta = 0.5$, $\log \dot{m} = -1$, $\log r_{tr} = 1.5$. Starting from below, the models correspond to $m = 10, 10^3, 10^5, 10^7, 10^9$. The upper two models are appropriate to AGNs.

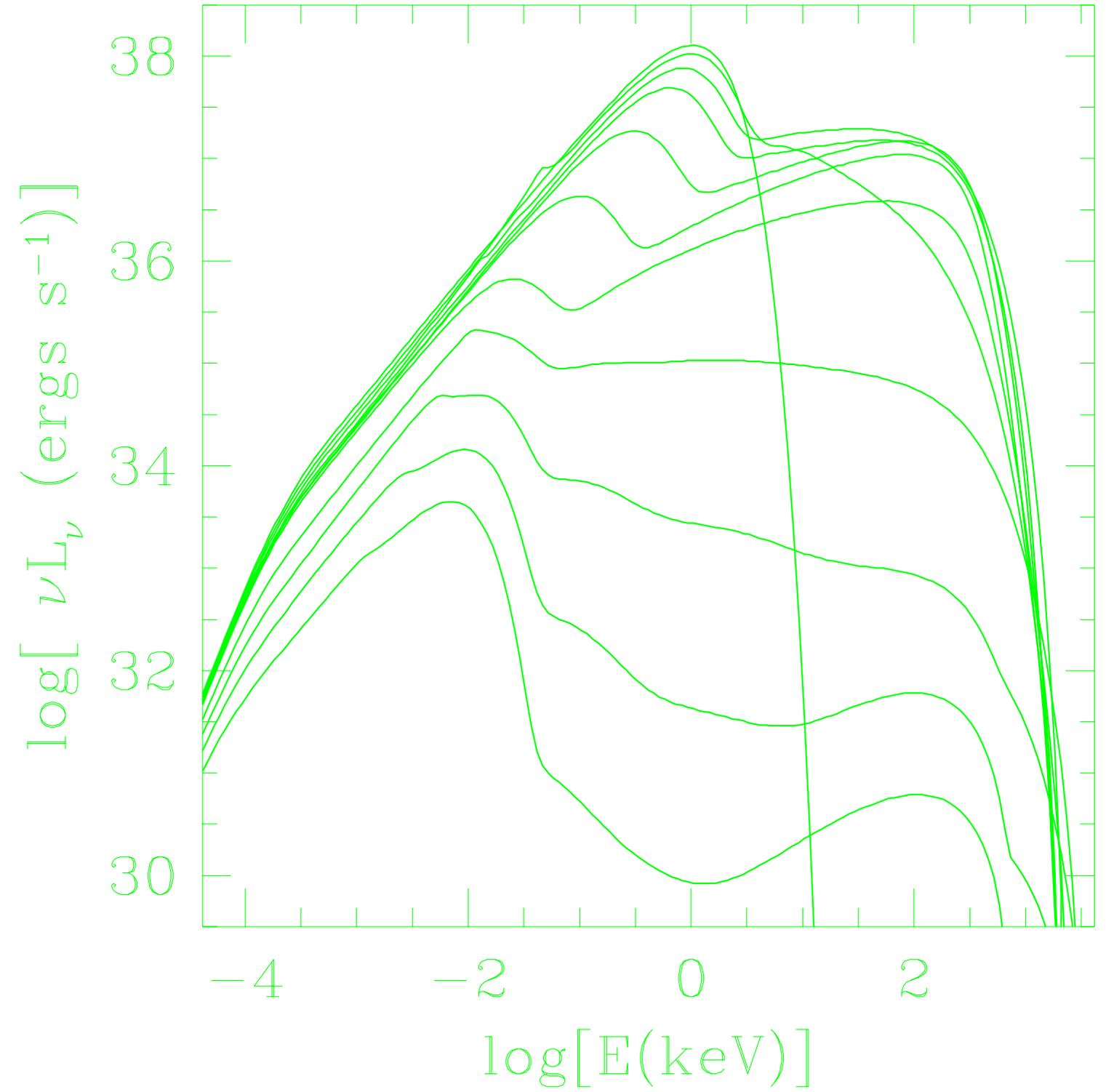

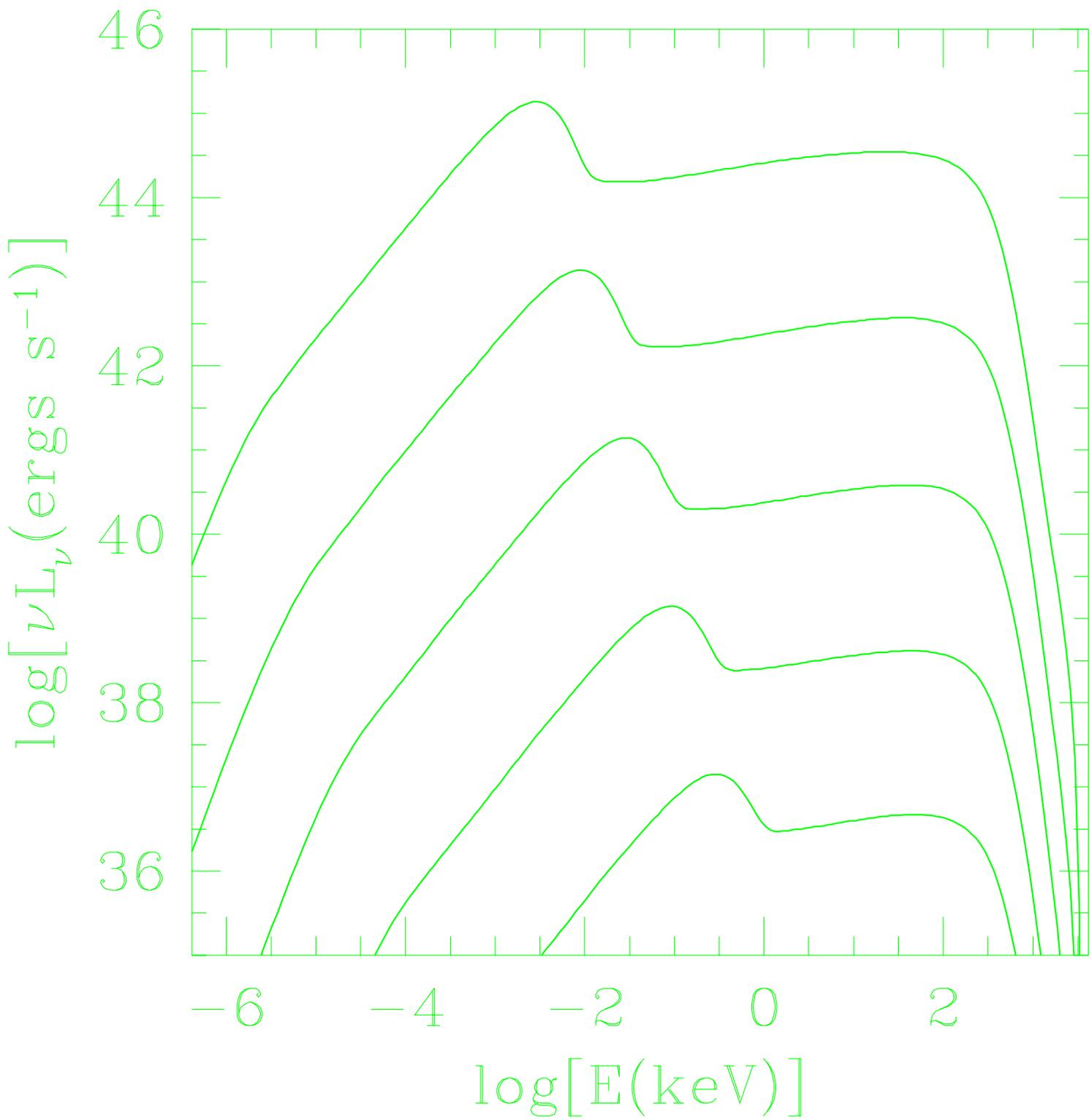